\documentclass[letterpaper,12pt]{article}

\pdfoutput=1  

\usepackage[top=1.0in,bottom=1.0in,left=1.0in,right=1.0in]{geometry}

\usepackage{amssymb}
\usepackage{amsmath}
\usepackage{url}

\newcommand\lhood{{\cal L}}
\newcommand\cls{CL{$_\textnormal{s}$}}

\begin{document}

\title{\bf \Large Comment on Frank Porter, ``Confidence intervals for the Poisson distribution"}
\author{Robert D. Cousins\thanks{cousins@physics.ucla.edu}\\
Dept.\ of Physics and Astronomy\\ 
University of California, Los Angeles\\
Los Angeles, California 90095}
 
\date{September 22, 2025}

\maketitle

\begin{abstract}
Frank Porter has recently posted a review of ``Confidence intervals for the Poisson distribution" (arXiv:2509.02852).  The long, diverse history of such intervals is closely related to that of confidence intervals for the parameter of the binomial distribution. While much of Porter's paper is enlightening and food for thought, I believe that his discussion of the intervals advocated by Gary Feldman and myself (FC) based on the likelihood ratio test (arXiv:physics/9711021) is flawed.  The fundamental point of disagreement is whether or not the likelihood function exists in a part of parameter space where the statistical model does not exist. (The new paper says yes; FC say no.) Here, I focus mainly on that issue and the consequences, along with a few other remarks.
\end{abstract}


\section{Introduction}
Frank Porter has recently posted a review of ``Confidence intervals for the Poisson distribution" \cite{porter2025} (referred to here as ``the paper").  The long, diverse history of such intervals is closely related to that of confidence intervals for the parameter of the binomial distribution, which we reviewed in 2010 \cite{cht2010}.\footnote{While not as detailed, our paper has more to say about Lancaster mid-P intervals, because of their use for the ratio of Poisson means problem, and also more about Wilson score intervals and the problems with Wald intervals ($\sqrt{N}$ intervals in the Poisson case).}  While much of the paper is enlightening and food for thought, I believe that his discussion of the intervals advocated by Gary Feldman and myself (FC) based on the likelihood ratio test~\cite{feldman1998} is flawed.  The fundamental point of disagreement is whether or not the likelihood function exists in a part of parameter space where the statistical model does not exist. (The paper says yes; FC say no.) Here, I focus mainly on that issue and the consequences, along with a few other remarks. 

We follow the notation of the paper, with sampling of a random variable $N=0,1,2,\ldots$ from a Poisson distribution of the form,
\begin{equation}
f(n;\theta,b) = \frac{\mu^n}{n!}e^{-\mu}=\frac{(\theta+b)^n}{n!}e^{-\theta-b},
\label{eq:Poisson}
\end{equation}
where $n$ is a possible value of $N$, $\mu$ is the mean of the Poisson distribution, equal to the sum of the ``signal strength'' $\theta$, considered to be the unknown quantity of interest, and a background strength $b\ge0$,
assumed here to be known and for the discussion here, independent of $\theta$. 

Eq.~\ref{eq:Poisson} is referred to by statisticians as a {\em statistical model} (or simply a model).  I emphasize that for the model to be completely defined, the domain of all parameters and observables must be specified.  In the cases under discussion (i.e., not a situation with destructive interference), $\theta\in[0,\infty)$ in Eq.~\ref{eq:Poisson}.

Historically (and evidently continuing), there has been confusion when it was not understood that the constraint $\theta\ge0$ is {\em part of the definition of the model}. The model {\em does not exist} for negative values of $\theta$.  It may appear that the math in Eq.~\ref{eq:Poisson} is fine as long as $\theta+b>0$, but we must keep in mind where Eq.~\ref{eq:Poisson} comes from.  In the usual situation, the signal and background events are {\em independent} Poisson processes, where the number of signal events $n_s$ is a sample from  
\begin{equation}
f_s(n_s;\theta) = \frac{\theta^{n_{\scriptstyle s}}}{n_s!}e^{-\theta},
\label{eq:signal}
\end{equation}
and the number of background events $n_b$ is a sample from 
\begin{equation}
f_b(n_b;b) = \frac{b^{n_{\scriptstyle b}}}{n_b!}e^{-b}.
\label{eq:background}
\end{equation}
Then there is a well-known theorem that says that these two independent Poisson processes combine to yield a total number of events $n$ according to Eq.~\ref{eq:Poisson}. However, this only makes sense if Eqs.~\ref{eq:signal} and \ref{eq:background} make sense. Thus, a fundamental property of the model is that $\theta>0$ (and of course $b>0$).

For observed $n<b$, it is fine to talk about (and use, and importantly report!) negative values of $n-b$, {\em since they exist in the model}. The problems arise when one uses values of $\theta$ that do not exist in the model. This key point, along with the analogous case of measuring non-negative masses with Gaussian resolution, is essential for understanding FC~\cite{feldman1998}, and is further discussed in my lectures~\cite{cousins2024lectures}.

 \section{Estimation conditional on the model being true}

Suppose that a particular experiment observes the number $n$. As in Eq.~2 of the paper, the
likelihood function of $\theta$ is obtained by inserting $n$ into Eq.~\ref{eq:Poisson} and considering it as a function  of $\theta$:
\begin{equation}
\label{lhood}
\lhood(\theta) = \frac{(\theta+b)^n}{n!}e^{-\theta-b},
\end{equation}
(Here I depart slightly from the paper's notation, but omit discussion of that.) In view of the above statements about Eq.~\ref{eq:Poisson}, we see that {\em the likelihood function does not exist for $\theta<0$}.  This non-existence has {\em nothing to do with a Bayesian prior for $\theta$}, or any post-data inference. It follows from the definition of the model.

Perhaps a word about models is in order.  While it is presumably true that models for real experiments are never perfect (and are all ``wrong" in the famous words of George Box), an enormous amount of effort often goes into constructing and validating models in particle physics, including adding carefully studied nuisance parameters to describe model uncertainty.  Thus, estimation in particle physics is typically performed conditional on the model (or one of a set of models) being true.

Certainly in the context of this discussion of the simplified model of Eq.~\ref{eq:Poisson}, we are assuming that the model is true. Then, what is the maximum likelihood estimate (MLE) of $\theta$
for observed $n<b$?  The MLE is at the boundary of the domain of $\theta$, i.e., $\theta=0$, {\em not} $\theta=n-b$ as claimed in the paper (e.g., in Section 1.1).\footnote{The subscript ``best" is used in Ref.~\cite{feldman1998}.}

Of course, one can take the mathematical expression in Eq.~\ref{lhood} and substitute negative values of $\theta$ into it.  The mistake is to continue to call it the likelihood function when doing so.  I do not know of any case in the statistics literature where an expression is called a likelihood function when it is not derived from a model with parameter values existing in the model.

\section{Discussion of Feldman-Cousins}
Section 8.2 says that FC has ``an additional condition designed to prevent intervals overlapping a `non-physical' region."
Section 10.3 says, ``The FC intervals also may not include the MLE. In this case, it is a design feature that $\theta<0$ is excluded and the MLE is redefined to never be quoted below zero." The true situation regarding ``redefining" is the opposite: 

\begin{enumerate}\item FC does not add a condition or redefine the MLE; it uses the correct MLE conditional on the model being true.  The FC intervals {\em always} contain the MLE because the likelihood ratio that determines the ordering of observations in the acceptance regions gives top ranking to the observation(s) that lead to the MLE.
\item It is the intervals in the paper that are obtained by a ``redefinition" of the likelihood function to be evaluated at values of $\theta$ that do not exist in the model (charitably called ``unphysical"). The intervals proposed by the paper, i.e., Garwood intervals with $b$ subtracted off the endpoints (Sections 3.12 and 4), do not contain the MLE when $b$ is large enough that the entire interval has negative values of $\theta$ and the MLE is 0.  
\end{enumerate}

I hasten to add that I have always agreed with the reporting requirements advocated by James and Roos~\cite{james1991}, favorably cited by the paper.  Papers should report the model, including the value of $b$, and the observed value of $n$, not just the derived quantities such as interval endpoints.\footnote{Modern software tools increasingly make it possible to report more complicated models.}  My objection is to referring to a value of $\theta$ not in the model as the ``measured value" (in the words of Ref.~\cite{james1991}) or the point estimate. The observed value of $n$ (or $n-b$) should just be called the observed (or sampled) value.

It is perhaps useful to have in mind the other prototypical example, discussed by Ref.~\cite{james1991}, namely the measurement of a physically positive quantity (neutrino mass-squared) with Gaussian resolution $\sigma$, i.e., with pdf \begin{equation}
\label{eqn-gaussian} 
p(x| \mu,\sigma) = 
   \frac{1}{\sqrt{2\pi\sigma^2}} \textnormal{e}^{-(x-\mu)^2/2\sigma^2},
\end{equation} 
where $\mu$ is the unknown true mass-squared and $x$ is the sampled value from the Gaussian. While the domain of $\mu$ is $[0,\infty)$, the domain of $x$ is $(-\infty,\infty)$. There is {\em nothing anomalous} about negative $x$ (!), which can occur up to half the time when $\mu\ll\sigma$. Of course, the sampled value of $x$ should be reported.  Again, it is incorrect to regard $x<0$ as the MLE. The MLE is 0 for $x<0$, since the model and the likelihood function do not exist for $\mu<0$. I discuss this case in more detail in Section 6.9 of my lectures~\cite{cousins2024lectures},

Section 8 of the paper is entitled ``Methods from particle physics", and discusses \cls\ and FC.\footnote{The notation for $p$-values in Eqs.~76 and 82 of the paper (and in its Ref.~38) follows the nonstandard convention for tail directions in the original \cls\ papers of Read.  In the standard convention for $p$-values, the relevant tail of one hypothesis is in the direction of the other hypothesis, as in the PDG RPP, Ref. [32] of the paper. In the standard convention, the ratio in Eq.~82 is $p_1/(1-p_0)$.}  While \cls\ does not correspond to any upper limit that has been found in the statistics literature (food for thought), the intervals in FC were soon understood (in time to add a note in proof) to be the confidence intervals that were obtained by ``inverting"~\cite{cousins2024lectures} the likelihood ratio test as described, for example, in ``Kendall and Stuart"~\cite{kendall1999}. The paper recognizes this for the case of $b=0$, but fails to see this for $b>0$, because of the disagreement described above regarding non-existence of the likelihood function for ``unphysical" values of $\theta$.

\section{Statistical Inference}
The paper's narrow use of the word ``inference" (essentially identified only with Bayesian statistics) risks confusion of its own, as the word is common in the frequentist statistics literature (albeit with historical controversy among Fisher, Neyman, Pearson, and many others since). For example, Volume 2A of ``Kendall and Stuart" \cite{kendall1999} is entitled {\em Classical Inference and the Linear Model} (from the era before the word ``frequentist" essentially replaced ``classical"). A quick search at Amazon on ``frequentist inference" finds that two of the first three hits are {\em Handbook of Bayesian, Fiducial, and Frequentist Inference}, edited by Berger, Meng, Reid, and Xie; and {\em Statistical Inference} by Migon, Gamerman,  and Louzada, described as ``an account of the Bayesian and frequentist approaches to statistical inference". I think that this is likely a case, as with ``estimation", where use of a word in everyday English is different from (varying) technical use in the statistics literature. 

Specifically, the use of ``inference" is troublesome in the paragraph in Section 1 that begins, ``On the other hand, we here wish to describe the result of the measurement, without making any further inference about the value of the parameter." Also, at the beginning of Section 1.4, we find, ``It is crucial to understand that our discussion is in the context of frequency statistics. That is, we are only interested here in providing a description of a measurement. There is no attempt to make a statement about the true mean of the Poisson distribution. We suggest that we are providing a description that could be useful input towards forming such an inference, but we stop short of actually making an inference. We belabor this point because it is neglected in many discussions, contributing to ongoing confusion."

 Equating ``frequentist statistics" with ``stopping short of actually making an inference" goes against the widespread use of ``statistical inference". If the point is to remind the reader that probabilistic statements about the true value of the parameter (or the truth of the model) are the domain of subjective Bayesian statistics, then that point can be made while carefully explaining the ways that ``inference" is used in the vast statistics literature. I think that this would be useful, and more successful than insisting that ``frequentist inference" not be used.\footnote{If I could change nearly a century of statistical jargon, I would ban ``noninformative prior" and change ``confidence interval" to ``coverage interval".}

\section{Concluding remarks}
Finally, the paper's emphasis on ``descriptive" intervals along with the insistence that ``inference" is not part of the frequentist paradigm, raises the question of how the proposed intervals would be regarded in practice. The conclusions of much of the experimental particle physics literature contain either an upper limit or a $p$-value (often expressed as a number of ``sigma" for significance of an effect). All the comparisons in the paper of lengths of intervals include the unphysical values in the various lengths.  But, it would seem to me that users interested in upper limits would (understandably) focus on the physically sensible part of the interval, viewing the upper endpoint as an upper limit, especially in the case of Garwood intervals, which are ``symmetric in probability" (Sections 3.8 and 5.1).

 I can imagine that one would in effect mentally make a ``cut" at 0, as in the paper's ``cut LR" intervals of Section 8.2. One can contemplate observations $n$  where the set of $\theta$ values that are in both the domain of the model and the confidence interval is the empty set. One would essentially be back to the 1980s, when people resorted to Bayesian intervals.\footnote{In fact, I encourage Bayesian intervals to be reported, but not exclusively~\cite{cousins2024lectures}.}  In that context, the lengths of FC intervals (as in Fig. 16) would not be so anomalous. In fact, they would continue to perform their original motivating feature of the absence of empty-set intervals (in the language of the paper, completely unphysical intervals).

\section*{Acknowledgments}
This work was partially supported by the U.S.\ Department of Energy under Award Number {DE}--{SC}0009937.


\begin{thebibliography}{1}%
\makeatletter
\providecommand{\hrefCMSnoop }[0]{\@secondoftwo}%
\makeatother

\bibitem{porter2025}
\hrefCMSnoop {} {F.~C. Porter, ``Confidence intervals for the Poisson distribution'',} (2025). \href{http://www.arXiv.org/abs/2509.02852 [physics.data-an]}{\texttt{ arXiv:2509.02852 [physics.data-an]}}.

\bibitem{cht2010}
\hrefCMSnoop {} {R.~D. Cousins, K.~E. Hymes, and J.~Tucker, ``{Frequentist evaluation of intervals estimated for a binomial parameter and for the ratio of Poisson means}'',} \textit{ Nucl. Instrum. Meth. A} \textbf{ 612} (2010) 388, \href{http://www.arXiv.org/abs/0905.3831}{\texttt{ arXiv:0905.3831}}. \url{http://arxiv.org/abs/0905.3831}. \href{http://dx.doi.org/10.1016/j.nima.2009.10.156}{\texttt{ doi:10.1016/j.nima.2009.10.156}}.

\bibitem{feldman1998}
\hrefCMSnoop {} {G.~J. Feldman and R.~D. Cousins, ``Unified Approach to the Classical Statistical Analysis of Small Signals'',} \textit{ Phys. Rev. D} \textbf{ 57} (1998) 3873, \href{http://www.arXiv.org/abs/physics/9711021}{\texttt{ arXiv:physics/9711021}}. \href{http://dx.doi.org/10.1103/PhysRevD.57.3873}{\texttt{ doi:10.1103/PhysRevD.57.3873}}.

\bibitem{cousins2024lectures}
\hrefCMSnoop {} {R.~D. Cousins, ``Lectures on Statistics in Theory: Prelude to Statistics in Practice'',} (2024). \href{http://www.arXiv.org/abs/1807.05996}{\texttt{ arXiv:1807.05996}}.

\bibitem{james1991}
\hrefCMSnoop {} {F.~James and M.~Roos, ``Statistical notes on the problem of experimental observations near an unphysical region'',} \textit{ Phys. Rev. D} \textbf{ 44} (1991) 299.

\bibitem{kendall1999}
A.~Stuart, K.~Ord, and S.~Arnold, ``Kendall's Advanced Theory of Statistics'', volume~2A.
\newblock Arnold, London, 6th edition, 1999.
\newblock See also earlier editions by Kendall and Stuart. The formal correspondence (duality) between hypothesis tests and confidence intervals is discussed in Chapter 20. The hypothesis test that is dual to the confidence intervals of Feldman and Cousins is detailed (including profile likelihood for nuisance parameters) in Chapter 22 on ``Likelihood Ratio Tests and Test Efficiency,'' pp.\ 238-239.

\end{thebibliography}

\providecommand{\href}[2]{#2}\begingroup\raggedright\endgroup

\end{document}